\documentclass{JHEP3}
 \usepackage{amssymb}

 \title{ Holographic Dual of Linear Dilaton Black Hole
 in Einstein-Maxwell-Dilaton-Axion Gravity }

 \author{Ran Li\\
 Institute of Modern Physics,\\
 Chinese Academy of Sciences,\\
  Lanzhou, 730000, Gansu, China\\
 \email{liran05@lzu.cn}}

 \author{ Ji-Rong Ren\\
 Institute of Theoretical Physics,\\
  Lanzhou University,\\
   Lanzhou, 730000, Gansu, China\\
 \email{renjr@lzu.edu.cn}}

 \abstract{
 Motivated by the recently
 proposed Kerr/CFT correspondence,
 we investigate the holographic dual
 of the extremal and non-extremal
 rotating linear dilaton black hole
 in Einstein-Maxwell-Dilaton-Axion Gravity.
 For the case of extremal black hole,
 by imposing the appropriate boundary
 condition at spatial infinity
 of the near horizon extremal geometry,
 the Virasoro algebra
 of conserved charges associated with the
 asymptotic symmetry group
 is obtained.
 It is shown that the microscopic entropy
 of the dual conformal field
 given by Cardy formula exactly agrees with
 Bekenstein-Hawking entropy of extremal black hole.
 Then, by rewriting the wave
 equation of massless scalar field
 with sufficient low energy
 as the SL(2, R)$_L$$\times$SL(2, R)$_R$ Casimir operator,
 we find the hidden conformal symmetry of
 the non-extremal linear dilaton black hole,
 which implies that the
 non-extremal rotating linear dilaton black hole
 is holographically dual to
 a two dimensional conformal field theory
 with the non-zero left and right temperatures.
 Furthermore, it is shown that the entropy of non-extremal
 black hole can be reproduced by using Cardy formula.
 }

 \date{\today}

 \newpage
 \begin{document}

 \section{Introduction}

 The recently proposed extremal
 Kerr/CFT correspondence \cite{GTSS}
 states that quantum gravity in the region
 very near the event horizon
 of an extreme Kerr black hole
 with proper boundary conditions
 is holographically dual to a two-dimensional
 chiral conformal field theory with the central
 charge proportional to angular momentum.
 The method employed by Guica,
 Hartman, Song and Strominger (GHSS)
 in \cite{GTSS} is very similar to the approach of
 Brown and Henneaux in \cite{Brown86},
 where the AdS$_3$ background
 is replaced by the near-horizon
 extremal Kerr (NHEK) geometry
 previously obtained in \cite{BH}.
 They shown that,
 by imposing the appropriate boundary
 condition at spatial infinity
 of the NHEK geometry,
 the conserved charges associated with the
 asymptotic symmetry group
 are found to form a copy of Virasoro algebra.
 If identifying this algebra with
 the Virasoro algebra of the dual two dimensional
 conformal field theory,
 it is shown that the macroscopic Bekenstein-Hawking
 entropy of extremal Kerr black hole can be reproduced
 by the microscopic entropy of dual conformal field theory
 via Cardy formula. This method has
 been generalized to calculate the
 entropies of extremal black holes in a
 lot of theories such as the
 Einstein theory, string theory, and
 supergravity theory. Some further
 studies on the extremal Kerr/CFT dual
 are listed in \cite{HMNS}-\cite{achilleas}.

 More recently, Castro, Maloney and Strominger (CMS)
 in a remarkable paper \cite{Castro} show that there exists a
 hidden SL(2, R)$_L\times$SL(2, R)$_R$ conformal symmetry for the
 four dimensional non-extremal Kerr black hole
 by studying the near-region wave equation of a
 massless scalar field.
 Interestingly, this hidden conformal symmetry is not derived from
 the conformal symmetry of spacetime geometry itself,
 but probed by the perturbation fields in the near region.
 It is shown that,
 for the massive scalar field
 in the background of Kerr black hole
 with sufficient low energy,
 the wave equation can be reproduced by the
 SL(2, R)$_L\times$SL(2, R)$_R$ Casimir operator.
 It is also shown that
 microscopic entropy
 computed by Cardy formula agrees exactly with
 the macroscopic Berenstein-Hawking
 entropy of non-extremal Kerr black hole.
 These observations suggest that non-extremal
 Kerr black hole is also holographically dual to a two-dimensional
 conformal field theory with non-zero left and right
 temperatures.
 Some related works on the hidden conformal
 symmetry of non-extremal black holes are
 listed in \cite{Krishnan}-\cite{ranliself}.

 Among the works on the Kerr/CFT dual,
 the asymptotic geometries of background spacetimes
 are of the flat or AdS type.
 In the present paper, we consider a
 rotating black hole
 in Einstein-Maxwell-Dilaton-Axion (EMDA) Gravity
 which is asymptotic to the linear dilaton
 spacetime.
 By using the approaches of Kerr/CFT dual,
 the holographic dual of the extremal
 and non-extremal rotating
 linear dilaton black hole
 is investigated.
 For the case of extremal black hole,
 the NHEK geometry is found after performing
 a coordinates transformation. This geometry
 is of the U(1)$_L\times$SL(2, R)$_R$ symmetry.
 By imposing the appropriate boundary
 condition at spatial infinity
 of the NHEK geometry, it is shown that
 the U(1)$_L$ symmetry is enhanced into
 a Virasoro algebra with the central charge
 $c=12J$, where $J$ is the angular momentum.
 One can conjecture that
 there exists a dual conformal field theory
 for the extremal linear dilaton black hole
 by identifying the Virasoro algebra with
 that of CFT.
 It is shown that the microscopic entropy
 of the dual conformal field
 given by Cardy formula exactly agrees with
 Bekenstein-Hawking entropy of extremal black hole.

 Then, by studying the wave
 equation of massless scalar field in this background,
 we study the hidden SL(2, R)$_L\times$SL(2, R)$_R$ conformal symmetry of
 the non-extremal black hole.
 We firstly find that the radial equation
 can be exactly solved by the hypergeometric functions.
 As hypergeometric functions
 transform in representations of SL(2,R), this suggests the
 existence of a hidden conformal symmetry.
 Furthermore, it is explicitly shown that the wave equation of scalar field
 can also be obtained by using of the SL(2, R)$_L\times$SL(2, R)$_R$ Casimir operator,
 which implies that the
 non-extremal rotating linear dilaton black hole
 is holographically dual to
 a two dimensional conformal field theory
 with the non-zero left and right temperatures.
 As a check of this conjecture, we also show that the entropy of non-extremal
 linear dialton black hole can be reproduced by using Cardy formula.

 This paper is organized as follows.
 In section II, we give a brief review
 of rotating linear dilaton black hole
 in Einstein-Maxwell-Dilaton-Axion Gravity.
 In section III,
 we obtain the near horizon geometry
 of extremal black hole and
 calculate the central charge and the left and right
 temperatures of the dual conformal field theory.
 We also find Bekenstein-Hawking entropy of extremal black hole
 matches with the microscopic entropy of dual CFT.
 In section IV, we study the hidden conformal
 symmetry of the non-extremal linear dilaton
 black hole by analysing the near-region wave
 equation of massless scalar field. Furthermore,
 the microscopic entropy of dual
 CFT with non-zero left and right
 temperatures are obtained.
 The last section is devoted to conclusion and discussion.

 \section{Rotating linear dilaton black hole}

 In this section, we will give a brief review about
 a class of rotating black hole solution
 of Einstein-Maxwell-dilaton-axion gravity in
 four dimensions with the linear
 dilaton background reported by Cl$\acute{\textrm{e}}$ment et al
 in \cite{clement}.
 The EMDA gravity theory can be considered arising as a
 truncated version of the bosonic sector of
 $D=4$, $N=4$ supergravity theory.
 The action of EMDA gravity theory
 is given by
 \begin{eqnarray}
 S=\frac{1}{16\pi}\int d^4 x\sqrt{-g}\left[
 -R+2\partial_\mu\phi\partial^\mu\phi
 +\frac{1}{2}e^{4\phi}\partial_\mu\kappa\partial^\mu\kappa
 -e^{-2\phi}F_{\mu\nu}F^{\mu\nu}
 -\kappa F_{\mu\nu}\tilde{F}^{\mu\nu}\right]\;,
 \end{eqnarray}
 where $\phi$ and $\kappa$ are the dilaton field and the axion (pseudoscalar)
 field respectively, $F$ and $\tilde{F}$ are the field strength of
 abelian vector field $A$ and its dual.
 Using the solution generating technique,
 the rotating black hole solution is obtained in
 \cite{clement}, where the metric without Nut charge is
 explicitly given by
 \begin{eqnarray}
 ds^2=-\frac{\Delta}{r_0r}dt^2+r_0r \left[\frac{dr^2}{\Delta}
 +d\theta^2+\textrm{sin}^2\theta\big(d\varphi-\frac{a}{r_0r}dt\big)^2 \right]\;\;,
 \end{eqnarray}
 where the factor $\Delta=(r^2-2Mr+a^2)$,
 and the other background fields are given by
 \begin{eqnarray}
 &&F=\frac{1}{\sqrt{2}}\left[\frac{r^2-a^2\textrm{cos}^2\theta}{r_0r^2}dr\wedge
 dt+a\textrm{sin}2\theta d\theta\wedge\big(d\varphi-\frac{a}{r_0r}dt\big)
 \right]\;\;, \nonumber\\
 &&e^{-2\phi}=\frac{r_0r}{r^2+a^2\textrm{cos}^2\theta}\;,\nonumber\\
 &&\kappa=-\frac{r_0a\textrm{cos}\theta}{r^2+a^2\textrm{cos}^2\theta}\;\;.
 \end{eqnarray}

 The metric (2.2) was derived from the Kerr metric
 by using the solution generating technique.
 When $r\rightarrow \infty$, the rotating
 linear dilaton black hole is asymptotic
 to the linear dilaton background, which
 is different with
 that of Kerr black hole.

 We now summarize the thermodynamics
 of rotating linear dilaton black hole.
 It should be noted that $M$ appeared in the solution is no longer the
 physical mass. In order to obtain the first law of black hole
 mechanics, the relevant thermodynamics
 quantities can be calculated by using the approach of Brown and
 York \cite{brown}. The mass $\widetilde{M}$ of rotating linear dilaton black hole
 is relative to the mass parameter $M$ towards the formula
 \begin{eqnarray}
 \widetilde{M}=\frac{M}{2}\;\;,
 \end{eqnarray}
 and the angular momentum $J$ is given by
 \begin{eqnarray}
 J=\frac{ar_0}{2}\;\;.
 \end{eqnarray}
 The Hawking temperature $T_H$, the Bekenstein-Hawking
 entropy $S_{BH}$ and the angular velocity
 $\Omega_H$ of the event horizon are  given as
 \begin{eqnarray}
 &&T_{H}=\frac{r_+-r_-}{4\pi r_0 r_+}\;\;,\nonumber\\
 &&\Omega_H=\frac{a}{r_0 r_+}\;\;,\nonumber\\
 &&S_{BH}=\pi r_0 r_+\;\;,
 \end{eqnarray}
 where $r_\pm=M\pm\sqrt{M^2-a^2}$ are the locations of
 the outer and the inner event horizon respectively.
 With the thermodynamical quantities given above,
 one can deduce the differential first law of black hole mechanics
 by straightforward calculation
 \begin{eqnarray}
 d\widetilde{M}=T_HdS_{BH}+\Omega_H dJ\;.
 \end{eqnarray}

 It should be noted that
 the electric charge $Q=r_0/\sqrt{2}$
 does not appear in the above thermodynamics relation.
 The differentiations
 are performed keeping $Q$ as a fixed value,
 which is characteristic of the linear dilaton background.
 When employing the approach of Brown and York to calculate the
 conserved quantities of the non-asymptotically flat spacetime, one
 should make a choice of the background 'vacuum' solution.
 In the present case, the electric charge $Q$ is selected to be
 the background charge. One can refer to \cite{clement} for more
 details.

 The extremality condition is $M=a$,
 and the entropy at extremality is
\begin{eqnarray}
 S_{BH}(T_H=0)=\pi r_0 a\;.
 \end{eqnarray}
 In the following two sections,
 we will try to reproduce the
 Bekenstein-Hawking entropies of the extremal and non-extremal
 linear dilaton black holes by using Cardy formula
 of the dual conformal field.

 \section{Holographic dual of extremal linear dilaton black hole }

 In this section, our purpose is to
 derive the Virasoro algebra of the dual
 conformal field
 by studying the asymptotic symmetry group
 of near horizon extremal geometry
 of linear dilaton black hole and
 reproduce its Bekenstein-Hawking entropy
 via Cardy formula.

 Firstly, we now try to explore the near-horizon geometry
 of extremal linear dilaton black hole. To do
 so, we need to perform the following coordinate transformations
 \begin{eqnarray}
 &&r=a+\epsilon \lambda \hat{r}\;,\nonumber\\
 &&t=\frac{\lambda \hat{t}}{\epsilon}\;,\nonumber\\
 &&\varphi=\hat{\varphi}+\frac{1}{r_0}\frac{\lambda
 \hat{t}}{\epsilon}\;,
 \end{eqnarray}
 with the parameter $\lambda^2=r_0 a$.
 After taking the $\epsilon\rightarrow 0$ limit, one can obtain
 the near-horizon geometry for an extremal rotating linear dilaton black
 hole
 \begin{eqnarray}
 ds^2=r_0 a \left(-\hat{r}^2d\hat{t}^2+
  \frac{d\hat{r}^2}{\hat{r}^2}+d\theta^2\right)
  +r_0 a\sin^2\theta
  \left(d\hat{\varphi}+\hat{r}d\hat{t}\right)^2\;,
 \end{eqnarray}
 and the near-horizon gauge field,
 dilaton field and axion field
 \begin{eqnarray}
 &&F=\frac{a}{\sqrt{2}}\left[
 \sin^2\theta d\hat{r}\wedge
 d\hat{t}+\sin 2\theta d\theta
 \wedge(\hat{\varphi}+\hat{r}d\hat{t})\right]\;,
 \nonumber\\
 &&e^{-2\phi}=\frac{r_0}{a(1+\cos^2\theta)}\;,
 \nonumber\\
 &&\kappa=-\frac{r_0\cos\theta}{a(1+\cos^2\theta)}\;.
 \end{eqnarray}

 It is worth noting that, although
 the asymptotic behavior of linear
 dilaton black hole is
 is different with
 that of Kerr black hole,
 the near-horizon metric (3.2)
 and the near-horizon background fields of an
 extremal linear dilaton black hole
 take the same form as that of
 an extremal Kerr black hole.

 The NHEK geometry (3.2) has an enhanced
 U(1)$_L\times$SL(2, R)$_R$ isometry group,
 which are respectively generated by the Killing
 vectors
 \begin{eqnarray}
 K_1=\partial_{\hat{\varphi}}\;,
 \end{eqnarray}
 and
  \begin{eqnarray}
 \bar{K}_1&=&\partial_{\hat{t}}\;,\nonumber\\
 \bar{K}_2&=&\hat{t}\partial_{\hat{t}}-\hat{r}\partial_{\hat{r}}\;,\nonumber\\
 \bar{K}_3&=&\left(\frac{1}{2\hat{r}^2}+\frac{\hat{t}^2}{2}\right)\partial_{\hat{t}}
 -\hat{t}\hat{r}\partial_{\hat{r}}-\frac{1}{\hat{r}}\partial_{\hat{\varphi}}\;.
 \end{eqnarray}

 We now employ the approach of Brown and Henneaux to
 find the central charge of the holographic dual
 conformal field theory description
 of an extremal rotating linear dilaton black hole.
 Because the linear dilaton black hole is a solution of
 EMDA gravity theory,
 it seems that there exists the non-vanishing
 contributions to the central charge from
 gauge field, dilaton field and axion field.
 Fortunately, an explicit calculation
 given by Compere et al. in \cite{compere} shows that,
 in Einstein-Maxwell-Dilaton theory
 with topological terms in four and five dimensions,
 the central charge receives no contribution from the
 non-gravitational fields, i.e.
 only the Einstein-Hilbert Lagrangian contributes
 to the value of the central charge.
 To find the
 central charge of the dual conformal field for
 the rotating linear dilaton black hole, for simplicity,
 it is sufficient to only calculate the gravitational
 field contribution.

 It is important to impose the appropriate boundary conditions at spatial
 infinity of NHEK geometry (3.2) and find the
 asymptotical symmetry group that preserves
 these boundary conditions.
 For the metric fluctuations around the NHEK geometry,
 we impose the boundary conditions
 \begin{eqnarray}
 \left(
   \begin{array}{cccc}
     h_{\hat{t}\hat{t}}=\mathcal{O}(\hat{r}^2)
     & h_{\hat{t}\hat{\varphi}}=\mathcal{O}(1)
     & h_{\hat{t}\theta}=\mathcal{O}(\frac{1}{\hat{r}})
     & h_{\hat{t}\hat{r}}=\mathcal{O}(\frac{1}{\hat{r}^2}) \\
      & h_{\hat{\varphi}\hat{\varphi}}=\mathcal{O}(1)
      & h_{\hat{\varphi}\theta}=\mathcal{O}(\frac{1}{\hat{r}})
      & h_{\hat{\varphi}\hat{r}}=\mathcal{O}(\frac{1}{\hat{r}}) \\
      &  & h_{\theta\theta}=\mathcal{O}(\frac{1}{\hat{r}})
      & h_{\theta\hat{r}}=\mathcal{O}(\frac{1}{\hat{r}^2})  \\
      &  &  & h_{\hat{r}\hat{r}}=\mathcal{O}(\frac{1}{\hat{r}^3}) \\
   \end{array}
 \right)
 \end{eqnarray}
 where $h_{\mu\nu}$ is the metric deviation
 from the near horizon geometry.

 The most general diffeomorphism symmetry that
 preserves such a boundary condition is generated by the
 vector field
 \begin{eqnarray}
 \zeta=\epsilon(\hat{\varphi})\frac{\partial}{\partial\hat{\varphi}}
 -\hat{r}\epsilon'(\hat{\varphi})\frac{\partial}{\partial\hat{r}}\;,
 \end{eqnarray}
 where $\epsilon(\hat{\varphi})$ is an arbitrary
 smooth periodic function of the coordinate $\hat{\varphi}$.
 It is convenient to define $\epsilon_n(\hat{\varphi})=-e^{-in\hat{\varphi}}$
 and $\zeta_n=\zeta(\epsilon_n)$, where $n$ are integers.
 Then the asymptotic symmetry group is generated by
 \begin{eqnarray}
 \zeta_n=-e^{-in\hat{\varphi}}\frac{\partial}{\partial\hat{\varphi}}
 -in\hat{r}e^{-in\hat{\varphi}}\frac{\partial}{\partial\hat{r}}\;,
 \end{eqnarray}
 which obey the Virasoro algebra
 with vanishing central charge
 \begin{eqnarray}
 i[\zeta_m,\zeta_n]=(m-n)\zeta_{m+n}\;.
 \end{eqnarray}

 Each diffeomorphism $\zeta_n$ is associated
 to a conserved charge defined by \cite{barnich}
 \begin{eqnarray}
 Q_{\zeta}=\frac{1}{8\pi}\int_{\partial\Sigma} k_{\zeta}\;,
 \end{eqnarray}
 where $\partial\Sigma$ is a spatial slice, and
 2-form $k_{\zeta}$ is defined as
 \begin{eqnarray}
 k_{\zeta}[h,g]&=&\frac{1}{2}\left[
 \zeta_\nu\nabla_\mu h-\zeta_\nu\nabla_\sigma h_\mu^{\;\;\sigma}
 +\zeta_\sigma\nabla_\nu h_\mu^{\;\;\sigma}
 +\frac{1}{2}h\nabla_\nu\zeta_\mu\right.\nonumber\\
 &&\left.-h_\nu^{\;\;\sigma}\nabla_\sigma\zeta_\mu
 +\frac{1}{2}h_{\nu\sigma}\left(\nabla_\mu\zeta^\sigma
 +\nabla^\sigma\zeta_\mu\right)
 \right]*\left(dx^\mu\wedge dx^\nu\right)\;,
 \end{eqnarray}
 where $*$ denotes the Hodge dual.

 The Dirac brackets of the conserved charges
 are just the common forms of the Virasoro algebras
 with a central term
 \begin{eqnarray}
 \left\{Q_{\zeta_{m}}, Q_{\zeta_{n}}\right\}_{D.B.}
 =Q_{[\zeta_{m}, \zeta_{n}]}+
 \frac{1}{8\pi}\int_{\partial\Sigma}
 k_{\zeta_m}[\mathcal{L}_{\zeta_n}g,g]\;,
 \end{eqnarray}
 Translated into the quantum version,
 the Virasoro algebra is given by
 \begin{eqnarray}
  \left[L_m, L_n\right]=(m-n)L_{m+n}+\frac{c}{12}(m^3+\alpha m)
  \delta_{m+n,0}\;,
 \end{eqnarray}
 where $c$ denote the central charge
 corresponding to the diffeomorphism
 and $\alpha$ is a trial constant.
 It follows that
 \begin{eqnarray}
 \frac{1}{8\pi}\int_{\partial\Sigma} k_{\zeta_m}[\mathcal{L}_{\zeta_n}g,g]
 =-\frac{i}{12}c(m^3+\alpha m)\delta_{m+n,0}\;,
 \end{eqnarray}
 Evaluating the integral for the case of
 the near-horizon metric of extremal linear dilaton black hole,
 we find the central charge
 \begin{eqnarray}
 c=6r_0 a\;.
 \end{eqnarray}

 It should be noted that
 the central charge can also be written
 as $c=12J$. This relation between
 the central charge and angular momentum
 is just of the same form as that for Kerr
 black hole in \cite{GTSS} and other examples
 of the Extremal Kerr/CFT dual.

 After obtaining the central charge of the extremal
 linear dilaton
 black hole, we now begin to get its CFT entropy. To get this, we have to
 calculate the generalized temperature with respect to the
 Frolov-Thorne vacuum.
 We consider the quantum field with
 eigenmodes of the asymptotic energy $\omega$ and angular momentum
 $m$, which are given by the following form
 \begin{eqnarray}
 e^{-i\omega t+im\varphi}=
 e^{-i\left(\omega-\frac{m}{r_0}\right)
 \frac{\lambda}{\epsilon}\hat{t}+im\hat{\varphi}}
 =e^{-in_R\hat{t}+in_L\hat{\varphi}}\;,
 \end{eqnarray}
 with
 \begin{eqnarray}
 n_R=\left(\omega-\frac{m}{r_0}\right)
 \frac{\lambda}{\epsilon}\;,\;\;\;n_L=m\;.
 \end{eqnarray}
 The correspondence Boltzmann factor is of the form
 \begin{eqnarray}
 e^{-\frac{\omega-m\Omega}{T_H}}=e^{-\frac{n_R}{T_R}-\frac{n_L}{T_L}}\;,
 \end{eqnarray}
 where the left and right temperatures are given by
 \begin{eqnarray}
 T_R&=&\frac{\lambda}{\epsilon}T_H\;,\nonumber\\
 T_L&=&\frac{T_H}{\frac{1}{r_0}-\Omega_H}\;.
 \end{eqnarray}
 In the extremal limit $M\rightarrow a$,
 the left and right temperatures reduce to
 \begin{eqnarray}
 T_R=0\;,\;\;\;T_L=\frac{1}{2\pi}\;.
 \end{eqnarray}

 According to the Cardy formula the entropy for a unitary CFT,
 we can obtain the microscopic entropy of the extremal linear
 dilaton black hole
 \begin{eqnarray}
 S_{CFT}=\frac{\pi^2}{3}cT_L=\pi r_0 a=S_{BH}(T_H=0)\;,
 \end{eqnarray}
 which precisely agrees with the Bekenstein-Hawking entropy.
 So one can conjecture that the extremal
 rotating linear dilaton black hole
 is dual to a two dimensional chiral
 conformal field theory.

 \section{Hidden conformal symmetry of non-extremal
 linear dilaton black hole }

 In this section, we investigate the hidden
 conformal symmetry of
 the non-extremal rotating linear dilaton balck hole.
 This hidden SL(2, R)$_L\times$SL(2, R)$_R$ conformal
 symmetry is not derived from the spacetime geometry
 itself, but can be probe by the perturbation field.
 Let us consider the wave equation of
 the neutral massless scalar field in the background
 of the non-extremal rotating linear dilaton black hole (2.2),
 which is given by the Klein-Gordon equation
 \begin{eqnarray}
 \frac{1}{\sqrt{-g}}\partial_\mu\left
 (\sqrt{-g}g^{\mu\nu}\partial_\nu\right)\Phi=0\;.
 \end{eqnarray}

 After performing the variable separation of scalar field
 $\Phi=e^{-i\omega t}R(r)P(\theta)e^{im\varphi}$,
 with $\omega$ and $m$ are quantum numbers,
 one can obtain the following
 two equations relative to radial part and angular part respectively
 \begin{eqnarray}
 &&\partial_r(\Delta\partial_r)R(r)+
 \left(\frac{(r_0r\omega-ma)^2}{\Delta}
 -\mathcal{K}^2\right)R(r)=0\;\;,\\
 &&\frac{1}{\textrm{sin}\theta}\partial_\theta
 (\textrm{sin}\theta\partial_\theta)P(\theta)-
 \left(\frac{m^2}{\textrm{sin}^2\theta}-\mathcal{K}^2
 \right)P(\theta)=0\;\;.
 \end{eqnarray}
 Unlike the case of Kerr black hole, now the angular equation is just
 that for the associated Legendre functions. So we have
 the separation constant
 \begin{eqnarray}
 \mathcal{K}^2=l(l+1)\;.
 \end{eqnarray}
 While the radial equation can be rewritten as
 \begin{eqnarray}
 \partial_r(\Delta\partial_r)R(r)+
 \frac{(\omega r_0 r_+-ma)^2}{(r-r_+)(r_+-r_-)}R(r)
 -\frac{(\omega r_0 r_--ma)^2}{(r-r_-)(r_+-r_-)}R(r)&&
 \nonumber\\
 =\left(l(l+1)-\omega^2 r_0^2\right)R(r)&&,
 \end{eqnarray}
 which can be exactly solved by the
 hypergeometric function without taking the near-horizon
 limit as we will show in the following.

 In order to solve the radial equation exactly,
 the new variable $z$ should be introduced like this
 \begin{eqnarray}
 z=\frac{r-r_+}{r-r_-}\;\;.
 \end{eqnarray}
 Then the radial equation (4.5) can be expressed as the following
 after some algebra
 \begin{eqnarray}
 z(1-z)\frac{d^2
 R}{dz^2}+(1-z)\frac{dR}{dz}+\left(\frac{A}{z}+\frac{B}{1-z}+C\right)R=0\;\;,
 \end{eqnarray}
 where $A$, $B$ and $C$ are given by
 \begin{eqnarray}
 A&=&\frac{(\omega r_0 r_+-ma)^2}{(r_+-r_-)^2}\;\;,\nonumber\\
 B&=&\omega^2 r_0^2-l(l+1)\;\;,\nonumber\\
 C&=&-\frac{(\omega r_0 r_--ma)^2}{(r_+-r_-)^2}\;\;.
 \end{eqnarray}
 By redefining the function $R(z)$ as
 \begin{eqnarray}
 R(z)=z^\alpha (1-z)^\beta F(z)\;\;,
 \end{eqnarray}
 with
 \begin{eqnarray}
 \alpha&=&-i\sqrt{A}=-i\frac{(\omega r_0 r_+-ma)}{r_+-r_-}\;\;,\nonumber\\
 \beta&=&\frac{1}{2}(1-\sqrt{1-4B})
 =\frac{1}{2}-\sqrt{(l+\frac{1}{2})^2-\omega^2r_0^2}\;\;,
 \end{eqnarray}
 the equation (4.7) can be transformed into the
 standard hypergeometric function form
 \begin{eqnarray}
 z(1-z)F''+(c-(1+a+b)z)F'-abF=0\;\;,
 \end{eqnarray}
 where the parameters $a$, $b$ and $c$ are given by
 \begin{eqnarray}
 a&=&\alpha+\beta+i\sqrt{-C}=\frac{1}{2}-
 \sqrt{(l+\frac{1}{2})^2-\omega^2 r_0^2}-i\omega r_0
 \;\;,\nonumber\\
 b&=&\alpha+\beta-i\sqrt{-C}=\frac{1}{2}-\sqrt{(l+\frac{1}{2})^2-\omega^2 r_0^2}
 -i\frac{\omega r_0(r_++r_-)-2ma}{r_+-r_-}
 \;\;,\nonumber\\
 c&=&1+2\alpha=1-\frac{2i(\omega r_0r_+-ma)}{r_+-r_-}\;\;.
 \end{eqnarray}
 The hypergeometric equation (4.11)
 has two linearly independent solutions which are given by
 \begin{eqnarray}
 f_1=F(a,b,c;z)\;\;,\;\;\;f_2=z^{(1-c)}F(a-c+1,b-c+1,2-c;z)\;\;,
 \end{eqnarray}
 where $F(a,b,c;z)$ is just the so-called hypergeometric function.
 Then, it follows that the general solution of the radial equation (4.5)
 can be expressed as
 \begin{eqnarray}
 R(z)=C_1 z^\alpha (1-z)^\beta F(a,b,c;z)+
 C_2 z^{-\alpha}(1-z)^\beta F(a-c+1,b-c+1,2-c;z)\;\;.
 \end{eqnarray}

 It should be noted that, generally,
 the wave equation cannot be analytically solved and the
 solution must be obtained by matching solutions in an overlap region
 between the near-horizon and asymptotic regions.
 But, in the present case, we have obtained the general
 solution (4.14) of wave equation and
 shown that the radial equation (4.5)
 can be exactly solved by hypergeometric functions.
 As hypergeometric functions
 transform in representations of SL(2, R), this suggests the
 existence of a hidden conformal symmetry.
 Now we will show that the radial equation
 can also be obtained by using of the SL(2, R) Casimir operator.

 Introducing the coordinates
 \begin{eqnarray}
 w^+&=&\sqrt{\frac{r-r_+}{r-r_-}}e^{2\pi T_R\varphi}\;,\nonumber\\
 w^-&=&\sqrt{\frac{r-r_+}{r-r_-}}e^{2\pi
 T_L\varphi+2n_L t}\;,\\
 y&=&\sqrt{\frac{r_+-r_-}{r-r_-}}e^{\pi(T_L+T_R)\varphi+n_L
 t}\;,\nonumber
 \end{eqnarray}
 with
 \begin{eqnarray}
 T_R=\frac{r_+-r_-}{4\pi a}\;,\;\;
 T_L=\frac{r_++r_-}{4\pi a}\;,\;\;
 n_L=\frac{1}{2r_0}\;.
 \end{eqnarray}
 Then we can locally define the vector fields
 \begin{eqnarray}
 H_1&=&i\partial_+\;,\nonumber\\
 H_0&=&i(w^+\partial_++\frac{1}{2}y\partial_y)\;,\\
 H_{-1}&=&i(w^{+2}\partial_++w^+y\partial_y-y^2\partial_-)\;,\nonumber
 \end{eqnarray}
 and
 \begin{eqnarray}
 \bar{H}_1&=&i\partial_-\;,\nonumber\\
 \bar{H}_0&=&i(w^-\partial_-+\frac{1}{2}y\partial_y)\;,\\
 \bar{H}_{-1}&=&i(w^{-2}\partial_-+w^-y\partial_y-y^2\partial_+)\;,\nonumber
 \end{eqnarray}
 These vector fields obey the SL(2, R) Lie algebra
 \begin{eqnarray}
 [H_0,H_{\pm 1}]=\mp iH_{\pm 1}\;,\;\;[H_{-1},H_1]=-2iH_0\;,
 \end{eqnarray}
 and similarly for $(\bar{H}_0,\bar{H}_{\pm1})$.
 The SL(2, R) quadratic Casimir operator is
 \begin{eqnarray}
 \mathcal{H}^2=\bar{\mathcal{H}}^2&=&-H_0^2+\frac{1}{2}(H_1H_{-1}+H_{-1}H_1)\nonumber\\
 &=&\frac{1}{4}(y^2\partial_y^2-y\partial_y)+y^2\partial_+\partial_-\;.
 \end{eqnarray}
 In terms of the $(t,r,\varphi)$ coordinates,
 the SL(2, R) generators are given by
 \begin{eqnarray}
 H_1&=&ie^{-2\pi T_R\varphi}\left[
 \sqrt{\Delta}\partial_r+\frac{1}{2\pi T_R}
 \frac{r-M}{\sqrt{\Delta}}\partial_\varphi
 -\frac{T_L}{T_R}\frac{r_0(Mr-a^2)}{M\sqrt{\Delta}}\partial_t
 \right]\;,\nonumber\\
 H_0&=&i\left[\frac{1}{2\pi T_R}\partial_\varphi
 -\frac{T_L}{T_R}r_0\partial_t\right]\;,\nonumber\\
 H_{-1}&=&ie^{2\pi T_R\varphi}\left[
 -\sqrt{\Delta}\partial_r+\frac{1}{2\pi T_R}
 \frac{r-M}{\sqrt{\Delta}}\partial_\varphi
 -\frac{T_L}{T_R}\frac{r_0(Mr-a^2)}{M\sqrt{\Delta}}\partial_t
 \right]\;,
 \end{eqnarray}
 and
 \begin{eqnarray}
 \bar{H}_1&=&ie^{-(2\pi T_L\varphi+\frac{t}{r_0})}\left[
 \sqrt{\Delta}\partial_r-\frac{1}{4\pi T_R}
 \frac{r_+-r_-}{\sqrt{\Delta}}\partial_\varphi
 +\frac{r_0 r}{\sqrt{\Delta}}\partial_t
 \right]\;,\nonumber\\
 \bar{H}_0&=&-ir_0\partial_t\;,\nonumber\\
 \bar{H}_{-1}&=&ie^{2\pi T_L\varphi+\frac{t}{r_0}}\left[
 -\sqrt{\Delta}\partial_r-\frac{1}{4\pi T_R}
 \frac{r_+-r_-}{\sqrt{\Delta}}\partial_\varphi
 +\frac{r_0 r}{\sqrt{\Delta}}\partial_t
 \right]\;,
 \end{eqnarray}
 and the SL(2, R) quadratic Casimir operator becomes
 \begin{eqnarray}
 \mathcal{H}^2=\partial_r\Delta\partial_r-
 \frac{(r_0 r_+\partial_t+a\partial_\varphi)^2}{(r-r_+)(r_+-r_-)}
 +\frac{(r_0 r_-\partial_t+a\partial_\varphi)^2}{(r-r_-)(r_+-r_-)}\;.
  \end{eqnarray}
 So for the scalar field with sufficient low energy $\omega r_0\ll
 1$, the near region wave equation can be written as
 \begin{eqnarray}
 \mathcal{H}^2\Phi=\bar{\mathcal{H}}^2\Phi=l(l+1)\Phi\;,
 \end{eqnarray}
 and the conformal weights of dual operator of the massless field $\Phi$ should be
 \begin{eqnarray}
 (h_L,h_R)=(l,l)\;.
 \end{eqnarray}

 So we have uncovered the hidden SL(2, R)$_L\times$SL(2, R)$_R$
 symmetry of the non-extremal rotating linear dilaton black hole.
 It should be noted that
 this symmetry is only locally defined and is spontaneously broken to
 U(1)$_L\times$U(1)$_R$ symmetry because of the periodic identification
 in the $\varphi$ coordinate.
 If one conjectures that the
 non-extremal rotating linear dilaton black hole
 is dual to a CFT,
 the broken of the conformal symmetry
 leads to the left temperature $T_L$ and right temperature $T_R$
 of the dual conformal field.

 As a check of this conjecture, we want to calculate the
 microscopic entropy of the dual CFT,
 and compare it with the Bekenstein-Hawking entropy
 of the non-extremal linear dilaton black hole.
 For the extremal case, the central charges can be
 derived from an analysis of the asymptotic
 symmetry group as we did in the last section.
 However, we did not know how to extend this
 calculation away from extremality. As did in \cite{Castro},
 we will simply assume that the
 conformal symmetry found here connects smoothly to
 that of the extreme limit and
 the central charge still keeps the same as the extremal case,
 which is given by  Eq.(3.15).
 The microscopic entropy of the dual CFT
 can be computed by the Cardy formula
 \begin{eqnarray}
 S_{CFT}=\frac{\pi^2}{3}(c_LT_L+c_RT_R)
 =\pi r_0 r_+=S_{BH}\;,
 \end{eqnarray}
 which matches with the black hole Bekenstein-Hawking entropy.

 \section{Conclusion}

 In this paper, we have extend the
 recently proposed Kerr/CFT correspondence
 to examine the dual conformal field
 of the extremal and non-extremal rotating linear dilaton black holes
 respectively.
 Firstly, for the extremal black hole,
 we have obtained its near horizon
 geometry and calculated
 the central charge and temperature of the dual conformal field
 by employing the approach of GHSS.
 It is shown that the microscopic entropy
 calculated by using Cardy formula agrees with the
 Bekenstein-Hawking entropy of the extremal
 black hole. Then, for the non-extremal case,
 we have investigated the hidden conformal symmetry of linear
 dilaton black hole by studying the wave
 equation of a massless scalar
 field, and found the
 left and right temperatures of the proposed dual
 conformal field.
 Furthermore, it is checked that
 the entropy of non-extremal
 linear dilaton black hole can also be reproduced
 by using Cardy formula.

 \section*{Acknowledgement}

 RL would like to thank Ming-Fan Li for
 helpful discussions.
 The work of JRR was supported by the Cuiying Programme of Lanzhou
 University (225000-582404) and the Fundamental Research Fund for
 Physics and Mathematic of Lanzhou University(LZULL200911).

 \end{document}